\documentclass{iau}
\usepackage{graphicx}

\def\hMSun{~ {h^{-1}\rm M_{\odot}}}
\newcommand{\MSun}{\mbox{${\rm M}_\odot$}}
\newcommand{\LCDM}{\mbox{$\Lambda$CDM}}
\def\hMpc{~ {h^{-1}\rm Mpc}}
\def\Mpc{~ {\rm Mpc}}

\def\parsec{~ {\rm parsec}}

\usepackage[authoryear]{natbib}

\title[The cosmic web in CosmoGrid voids]
{The cosmic web in CosmoGrid void regions}

\author[Rieder et al.]
{Steven~Rieder$^{1,2}$,
Rien~van~de~Weygaert$^{1}$,
Marius~Cautun$^{3,1}$,
Burcu~Beygu$^{1}$
\and Simon~Portegies~Zwart$^{2}$}

\affiliation{$^1$Kapteyn Instituut, Rijksuniversiteit Groningen, P.O. Box 800,
  9700 AV Groningen, \\The Netherlands, email: {\tt steven@stevenrieder.nl}\\
  $^2$ Sterrewacht Leiden, Leiden University, P.O. Box 9513, 2300 RA
  Leiden, The Netherlands,\\
  $^3$ Department of Physics, Institute for Computational Cosmology,
  University of Durham, South Road, Durham DH1 3LE, UK}

\pubyear{2014}
\volume{308} 
\jname{The Zeldovich Universe: Genesis and Growth of the Cosmic Web}
\editors{R.~van~de~Weygaert, S.~Shandarin, E.~Saar \& J.~Einasto, eds.}
\begin{document}

\maketitle

\begin{abstract}
We study the formation and evolution of the cosmic web, using the
high-resolution CosmoGrid \LCDM\ simulation. In particular, we
investigate the evolution of the large-scale structure around void
halo groups, and compare this to observations of the VGS-31 galaxy
group, which consists of three interacting galaxies inside a large
void.  

The structure around such haloes shows a great deal of tenuous
structure, with most of such systems being embedded in intra-void
filaments and walls. We use the {\tt Nexus+} algorithm to detect walls
and filaments in CosmoGrid, and find them to be present and detectable
at every scale. The void regions embed tenuous walls, which in turn
embed tenuous filaments. We hypothesize that the void galaxy group of
VGS-31 formed in such an environment. 

\end{abstract}

\section{Introduction}

The large-scale structure of the Universe is volume-dominated by
voids: enormous regions of space that contain very few galaxies
(\citealp[see][]{1982Natur.300..407Z}, and
\citealp{2011IJMPS...1...41V} for a review). They are surrounded by
walls, filaments and clusters, together forming the {\em Cosmic Web}
\citep{1996Natur.380..603B}. \cite{2011AJ....141....4K,
2012AJ....144...16K} conducted a survey of 60 void galaxies, selected
from the SDSS. One of these, VGS-31, was later revealed to be an
interacting galaxy system, consisting of three galaxies that appeared
to be aligned along an HI filament \citep{2013AJ....145..120B}. It is
a manifestation of void substructure as a result of hierarchical
evolution \citep{1993MNRAS.263..481V, 2004MNRAS.350..517S,
2010MNRAS.404L..89A, 2013MNRAS.428.3409A}.

\section{Simulation and methods} 

In order to investigate the formation of a system like VGS-31, we
study the environment of dark matter haloes in cosmic voids within the
high-resolution CosmoGrid \LCDM\ suite of simulations
\citep{2010IEEEC..43...63P, 2013ApJ...767..146I}. CosmoGrid was
performed using the GreeM/SUSHI code \citep{2009PASJ...61.1319I,
2011CS&D....4a5001G}. The most detailed realisation contains
$2048^3$ particles in a $(30\Mpc)^3$ volume, resulting in a mass
of $1.28\times10^5\MSun$ for each particle. With a softening length
$\epsilon$ of $175 \parsec$ at $z=0$ and over 500 snapshots, it has a
very high mass-, spatial- and time-resolution. 

Due to its high resolution, CosmoGrid is especially suitable for case
studies of smaller dark matter haloes (up to Milky Way and Local
Group-size). However, its limited volume makes it less suited for
studies requiring statistical representation.

For the analysis of the CosmoGrid environment, we use the {\tt Nexus+}
\citep{2013MNRAS.429.1286C,2014MNRAS.441.2923C} and {\tt DTFE}
\citep{2000A&A...363L..29S, 2009LNP...665..291V, 2011arXiv1105.0370C}
tools and the AMUSE \citep{2013CoPhC.183..456P, 2013A&A...557A..84P}
framework. {\tt DTFE} uses the Voronoi and Delaunay tessellations of
the particle distribution to construct a volume-weighted linear
piecewise density field.  The inverse volume of the contiguous Voronoi
cells provide a local density estimate,  and the Delaunay tessellation
are used as an interpolation grid. The resulting field reconstruction
retains the anisotropic character and multiscale structure of the
particle distribution. We use {\tt DTFE} to determine densities in the
volume (smoothed on a scale of $1\hMpc$), where we identify regions
with a density contrast $\delta < 0$ as void regions. 

\begin{figure}[htbp]
  \begin{center} 
    \includegraphics[width=0.43\textwidth]{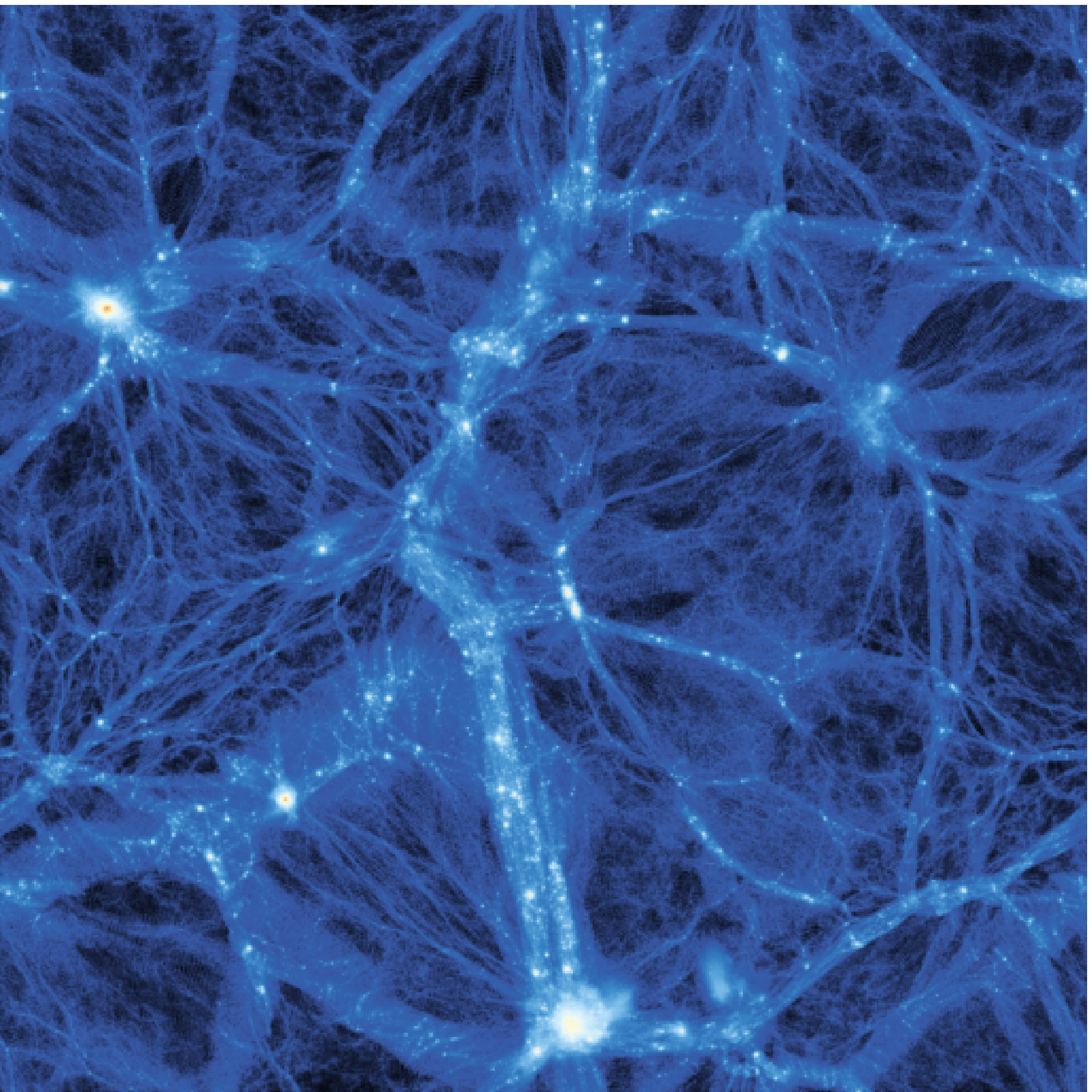}
    \includegraphics[width=0.43\textwidth]{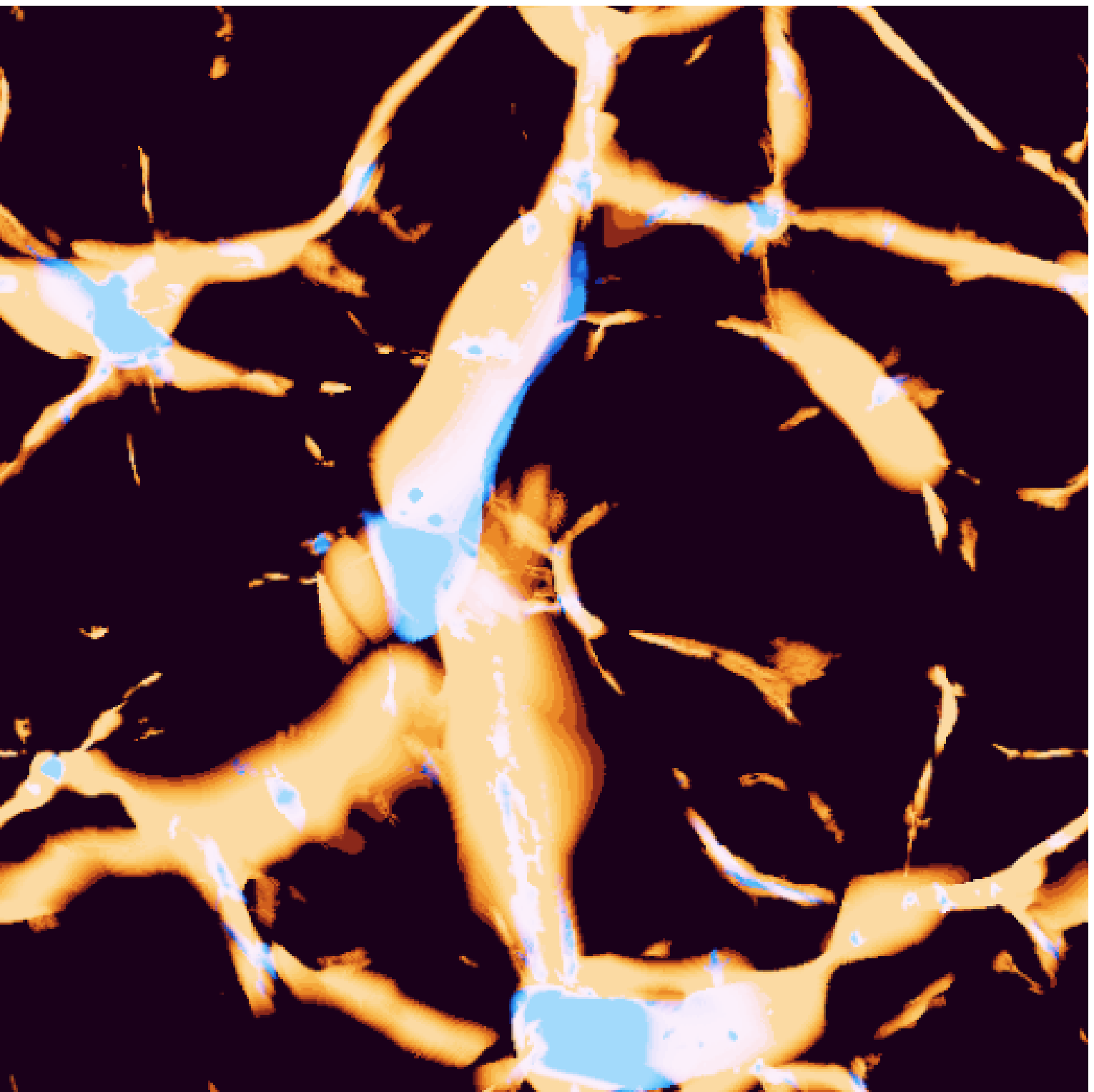}
    \caption{600kpc thick slice of the $1024^3$-particle CosmoGrid
      realization at z=0. Left: particles. Right: Nexus+ analysis of
      the same region. Blue regions represent filaments and clusters,
      while orange represents the walls. \label{fig:Nexus}}
  \end{center} 
\end{figure}

Additionally, we use {\tt Nexus+} to determine whether a region is
part of a wall, a filament or a cluster.  {\tt Nexus+} is a technique
for the multiscale morphological characterisation of the spatial mass
distribution. It is an extension and improvement of the Multiscale
Morphology Filter \citep{2007A&A...474..315A}, in which the locally
dominant morphological signal is extracted from the 4d scale-space
representation of the mass distribution.  It dissects the mass
distribution into voids, walls, filaments and cluster nodes on the
basis of the local signature of the Hessian of the logarithm of the
density field.  Using this method, we analyse the full volume of
CosmoGrid (see figure~\ref{fig:Nexus}). Due to the multi-scale nature
of the cosmic environment, a void region found with a $1\hMpc$
smoothing scale can still contain walls, filaments and clusters on a
smaller scale.

\section{VGS-31-like systems in CosmoGrid}

\subsection{Candidate systems}
We use CosmoGrid data to investigate possible formation scenarios for
the VGS-31 system, and to find out if the filamentary structure seen
in the alignment of these galaxies is to be expected or coincidental.
To this end, we first search for systems of similar properties to
VGS-31 in CosmoGrid. Using the mass, environmental density contrast
and size of the system as selection criteria, we find eight candidate
systems in CosmoGrid that we name ``CosmoGrid Void Systems''
\citep[CGV, see table~\ref{tab:cgv} and][for
details]{2013MNRAS.435..222R}.  While at $z=0$ these systems appear
very similar, their formation histories are quite diverse: some
systems continue to experience mergers up to $z=0$, while others
remain virtually unchanged since $z=6$.  Here, we discuss the
environment of these haloes.

\begin{table}[htbp]
  \footnotesize
  \begin{center}
  \begin{tabular}{llllcc}
    \hline
    Name  & $M_{\mbox{vir}} (\hMSun)$ & Last MM (Gyr)& $\delta$ &
    Wall & Filament\\
    \hline
    CGV-A & $3.15\times 10^{10}$ & -    & -0.68 & X & -\\
    CGV-B & $3.95\times 10^{10}$ & 5.24 & -0.51 & X & X\\
    CGV-C & $2.99\times 10^{10}$ & 1.19 & -0.51 & X & -\\
    CGV-D & $4.60\times 10^{10}$ & 10.9 & -0.63 & X & X\\
    CGV-E & $1.99\times 10^{10}$ & 2.44 & -0.57 & X & X\\
    CGV-F & $2.27\times 10^{10}$ & -    & -0.62 & X & X\\
    CGV-G & $2.14\times 10^{10}$ & 5.80 & -0.61 & X & X\\
    CGV-H & $4.63\times 10^{10}$ & 8.45 & -0.50 & X & -\\
    \hline
  \end{tabular}
  \caption{VGS-31-like systems in CosmoGrid. 1: Virial mass of the main halo,
    2: Time of the last major merger, 3: Density contrast, 4: System is
    embedded in a wall, 5: System is embedded in a filament.\label{tab:cgv}}
  \end{center}
\end{table}

\subsection{Environment of the systems}

\begin{figure}[htb]
  \begin{center}
    \includegraphics[width=0.85\textwidth]{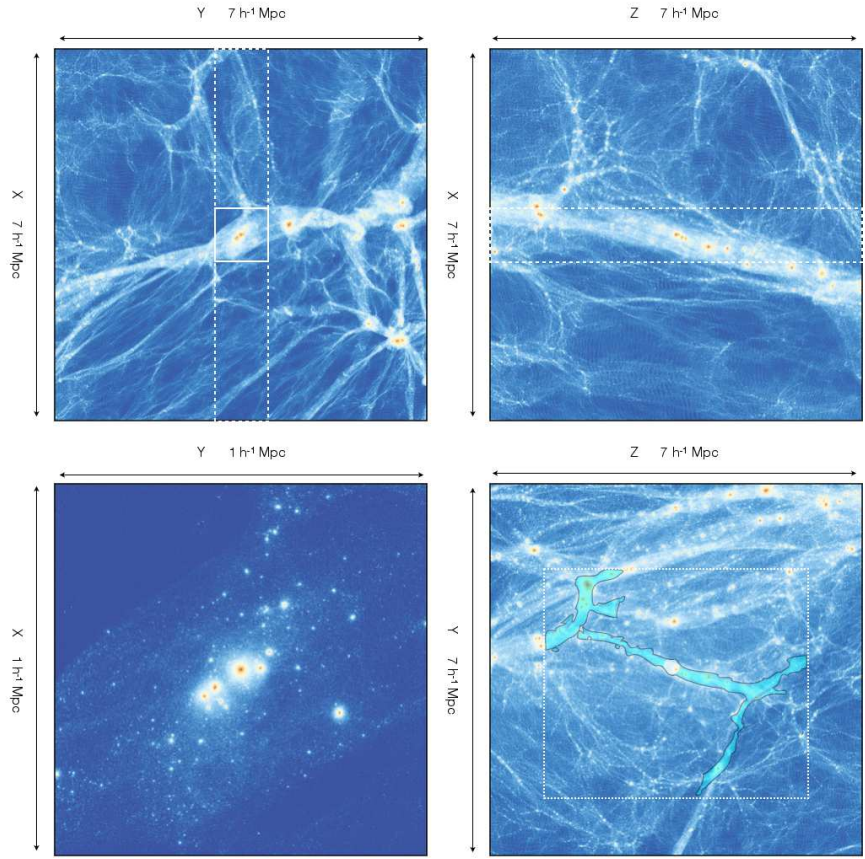}
    \caption{Environment of CGV-G, as seen from different angles, in
      $1\hMpc$ thick slices.  Lower-right panel includes a {\tt Nexus+}
      overlay. \label{fig:cgvg-environment}}
  \end{center}
\end{figure}

In figure~\ref{fig:cgvg-environment}, we show the environment of CGV-G
as a representative for the void halo systems, at a $7\hMpc$ scale.
While the whole system is in a void, its environment is highly
structured. The tenuous wall, which in turn contains a thin filament,
is clearly visible in this figure. 

While the majority of the CGV systems forms along a tenuous filament,
all of the systems are embedded in thin but prominent walls (see
table~\ref{tab:cgv}). These walls have a typical thickness of around
$0.4\hMpc$. They show a strong coherence and retain the character of a
highly flattened structure out to a distance of at least $3\hMpc$ from
the CGV haloes.  The filaments in which five of the systems reside are
rather short, not longer than $4\hMpc$, with a diameter of around
$0.4\hMpc$.  Compared to the high-density filaments of the
larger-scale cosmic web, the environment of void haloes is very
feeble.  At earlier times, this structure inside the void is much more
prominent, becoming more tenuous over time until it can hardly be
detected at $z=0$. However, it remains visible in the alignment of the
haloes. The formation histories of these haloes are quite diverse,
with the main haloes forming between $z=6$ and $z=0$.

\begin{figure}[htbp]
  \begin{center}
    \includegraphics[width=0.85\textwidth]{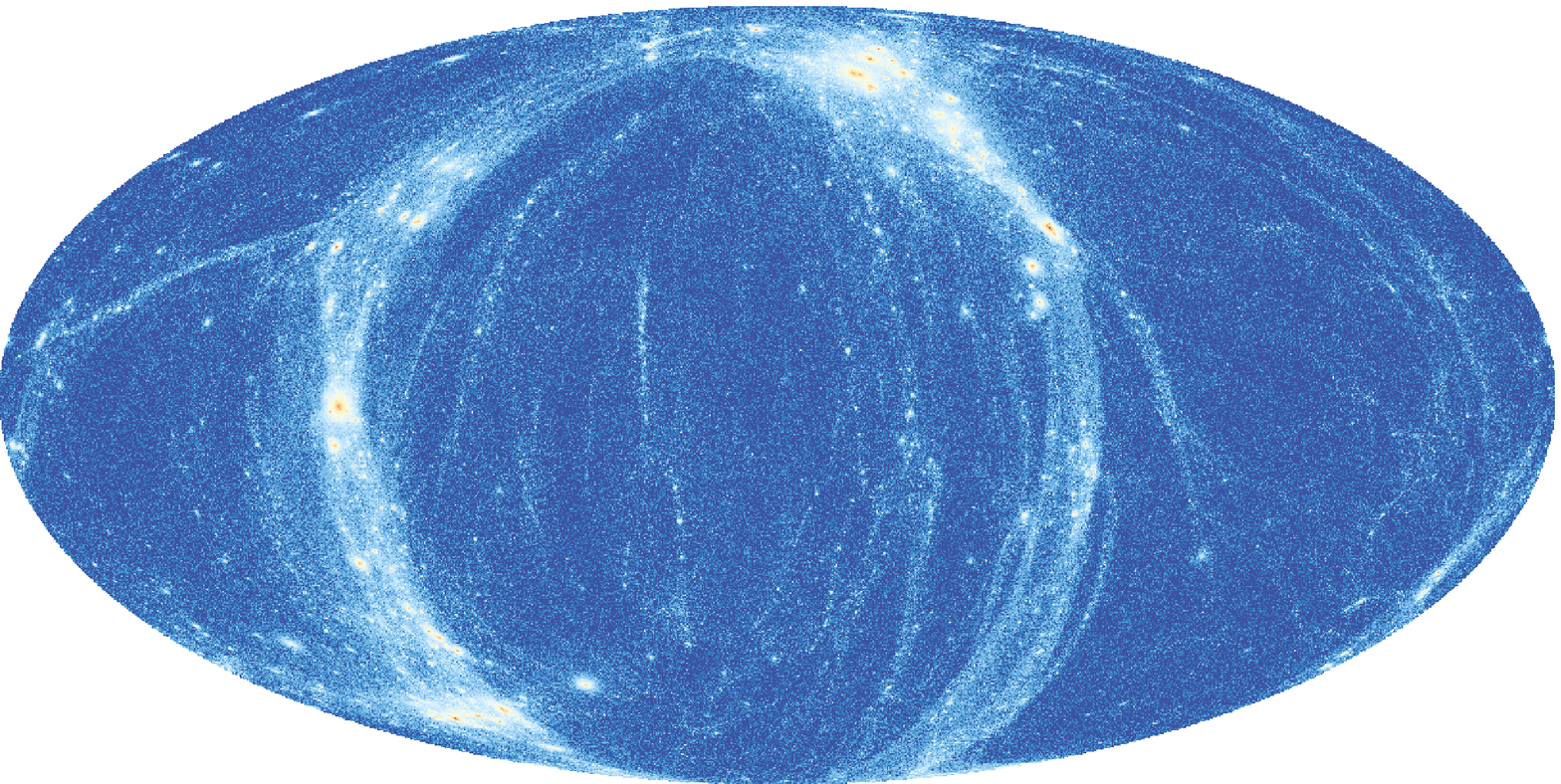}
    \caption{Mollweide projection of the environment around CGV-G at
      $z=0$, up to a distance of $2\hMpc$ from the main halo's centre.
      The wall manifests itself as a circle, while the filament is
      visible within the wall as two bright regions on opposite sides
      (top-right and lower-left).  \label{fig:cgvg-sky}}
  \end{center}
\end{figure}

In figure~\ref{fig:cgvg-sky}, we show a Mollweide projection of all
the dark matter around CGV-G up to a distance of $2\hMpc$.  This
projection clearly shows the wall in which the system is embedded,
while the filament appears as two bright spots on opposite sides. We
see very similar structure around the other CGV-systems.

A similar study within a larger volume will provide a more definitive
answer on the abundance of VGS-31-like systems. Furthermore, a
continuation study with a hydrodynamical simulation will provide
insight into the gaseous surroundings of such systems.

\section*{Acknowledgements}
The authors are grateful to Tomoaki~Ishiyama and Bernard~Jones, for
many insightful discussions. 
This work was supported by NWO grants IsFast [\#643.000.803], VICI
[\#639.073.803], LGM [\#612.071.503] and AMUSE [\#614.061.608]), NCF
(grants [\#SH-095-08] and [\#SH-187-10]), NOVA and the LKBF in the
Netherlands. 
SR and RvdW acknowledge support by the John Templeton Foundation,
grant [\#FP05136-O].

\newcommand{\aj}{AJ}
\newcommand{\actaa}{Acta Astron.}
\newcommand{\araa}{ARA\&A}
\newcommand{\apj}{ApJ}
\newcommand{\apjl}{ApJ}
\newcommand{\apjs}{ApJS}
\newcommand{\ao}{Appl.~Opt.}
\newcommand{\apss}{Ap\&SS}
\newcommand{\aap}{A\&A}
\newcommand{\aapr}{A\&A~Rev.}
\newcommand{\aaps}{A\&AS}
\newcommand{\azh}{AZh}
\newcommand{\baas}{BAAS}
\newcommand{\caa}{Chinese Astron. Astrophys.}
\newcommand{\cjaa}{Chinese J. Astron. Astrophys.}
\newcommand{\icarus}{Icarus}
\newcommand{\jcap}{J. Cosmology Astropart. Phys.}
\newcommand{\jrasc}{JRASC}
\newcommand{\memras}{MmRAS}
\newcommand{\mnras}{MNRAS}
\newcommand{\na}{New A}
\newcommand{\nar}{New A Rev.}
\newcommand{\pra}{Phys.~Rev.~A}
\newcommand{\prb}{Phys.~Rev.~B}
\newcommand{\prc}{Phys.~Rev.~C}
\newcommand{\prd}{Phys.~Rev.~D}
\newcommand{\pre}{Phys.~Rev.~E}
\newcommand{\prl}{Phys.~Rev.~Lett.}
\newcommand{\pasa}{PASA}
\newcommand{\pasp}{PASP}
\newcommand{\pasj}{PASJ}
\newcommand{\qjras}{QJRAS}
\newcommand{\rmxaa}{Rev. Mexicana Astron. Astrofis.}
\newcommand{\skytel}{S\&T}
\newcommand{\solphys}{Sol.~Phys.}
\newcommand{\sovast}{Soviet~Ast.}
\newcommand{\ssr}{Space~Sci.~Rev.}
\newcommand{\zap}{ZAp}
\newcommand{\nat}{Nature}
\newcommand{\iaucirc}{IAU~Circ.}
\newcommand{\aplett}{Astrophys.~Lett.}
\newcommand{\apspr}{Astrophys.~Space~Phys.~Res.}
\newcommand{\bain}{Bull.~Astron.~Inst.~Netherlands}
\newcommand{\fcp}{Fund.~Cosmic~Phys.}
\newcommand{\gca}{Geochim.~Cosmochim.~Acta}
\newcommand{\grl}{Geophys.~Res.~Lett.}
\newcommand{\jcp}{J.~Chem.~Phys.}
\newcommand{\jgr}{J.~Geophys.~Res.}
\newcommand{\jqsrt}{J.~Quant.~Spec.~Radiat.~Transf.}
\newcommand{\memsai}{Mem.~Soc.~Astron.~Italiana}
\newcommand{\nphysa}{Nucl.~Phys.~A}
\newcommand{\physrep}{Phys.~Rep.}
\newcommand{\physscr}{Phys.~Scr}
\newcommand{\planss}{Planet.~Space~Sci.}
\newcommand{\procspie}{Proc.~SPIE}
\bibliographystyle{plainnat}
\bibliography{references}

\end{document}